\documentclass[aps,prd,superscriptaddress,twocolumn,amsfonts,amssymb,amsmath,showpacs,10pt]{revtex4-2}

\usepackage{bm}
\usepackage{amsfonts}
\usepackage{amssymb}
\usepackage{amsmath}
\usepackage{latexsym}
\usepackage[latin1]{inputenc}
\usepackage[T1]{fontenc}

\usepackage{graphicx}
\usepackage[caption=false]{subfig} 
\usepackage{booktabs}
\usepackage{dcolumn}
\usepackage{multirow}
\usepackage{hhline}

\usepackage[british]{babel}
\usepackage{textcomp}
\usepackage{ragged2e}
\usepackage{commath}

\usepackage{xcolor}
\usepackage{orcidlink}
\usepackage{epsfig}

\def \nn  {\nonumber}

\def\jnl@style{\it}
\def\aaref@jnl#1{{\jnl@style#1}}

\def\aaref@jnl#1{{\jnl@style#1}}

\def\aj{\aaref@jnl{AJ}}                   
\def\apj{\aaref@jnl{ApJ}}                 
\def\apjl{\aaref@jnl{ApJ}}                
\def\apjs{\aaref@jnl{ApJS}}               
\def\apss{\aaref@jnl{Ap\&SS}}             
\def\aap{\aaref@jnl{A\&A}}                
\def\aapr{\aaref@jnl{A\&A~Rev.}}          
\def\aaps{\aaref@jnl{A\&AS}}              
\def\mnras{\aaref@jnl{Mon.~Not.~Roy.~Astron.~Soc.}}             
\def\prd{\aaref@jnl{Phys.~Rev.~D}}        
\def\prc{\aaref@jnl{Phys.~Rev.~C}}  
\def\prl{\aaref@jnl{Phys.~Rev.~Lett.}}    
\def\qjras{\aaref@jnl{QJRAS}}             
\def\skytel{\aaref@jnl{S\&T}}             
\def\ssr{\aaref@jnl{Space~Sci.~Rev.}}     
\def\zap{\aaref@jnl{ZAp}}                 
\def\nat{\aaref@jnl{Nature}}              
\def\aplett{\aaref@jnl{Astrophys.~Lett.}} 
\def\apspr{\aaref@jnl{Astrophys.~Space~Phys.~Res.}} 
\def\physrep{\aaref@jnl{Phys.~Rep.}}      
\def\physscr{\aaref@jnl{Phys.~Scr}}       
\def\commat{\aaref@jnl{Comm.~Math.~Phys.}}              
\def\science{\aaref@jnl{Science}}               
\def\cqg{\aaref@jnl{Classical Quant.~Grav.}}            
\def\jpcs{\aaref@jnl{JPCS}}                                     
\def\ijmpd{\aaref@jnl{Int.~J.~Mod.~Phys.~D}}                    
\def\grg{\aaref@jnl{Gen.~Relat.~Gravit.}}               
\def\rpp{\aaref@jnl{Rep.~Prog.~Phys.}}          
\def\npa{\aaref@jnl{Nucl.~Phys.~A}}        
\def\lrr{\aaref@jnl{Living Rev.~Rel.}}                   
\def\jcap{\aaref@jnl{J.~Cosmology Astropart.~Phys.}}    
\def\rmp{\aaref@jnl{Rev.~Mod.~Phys.}}   
\def\epjc{\aaref@jnl{Eur.~Phys.~J.~C}} 
\def\plb{\aaref@jnl{~Phy.~Lett.~B}} 
\def\mpla{\aaref@jnl{Mod.~Phy.~Lett.~A}} 
\def\arxiv{\aaref@jnl{arxiv.org}}

\allowdisplaybreaks[1]

\usepackage{hyperref}
\hypersetup{
    colorlinks=true,
    citecolor=blue,
    linkcolor=red,
    filecolor=magenta,
    urlcolor=cyan
}
\begin{document}

\color{black}  

\title{\bf Propagating gravitational waves in teleparallel Gauss-Bonnet gravity}

\author{Shivam Kumar Mishra \orcidlink{0009-0006-4754-4103
}}
\email{shivamkumarmishra.mt@gmail.com}
\affiliation{Department of Mathematics,
Birla Institute of Technology and Science-Pilani, Hyderabad Campus, Jawahar Nagar, Kapra Mandal, Medchal District, Telangana 500078, India.}

\author{Jackson Levi Said\orcidlink{0000-0002-7835-4365}}
\email{jackson.said@um.edu.mt}
\affiliation{Institute of Space Sciences and Astronomy, University of Malta, Malta, MSD 2080}
\affiliation{Department of Physics, University of Malta, Malta}

 \author{B. Mishra\orcidlink{0000-0001-5527-3565}}
 \email{bivu@hyderabad.bits-pilani.ac.in}
 \affiliation{Department of Mathematics,
Birla Institute of Technology and Science-Pilani, Hyderabad Campus, Jawahar Nagar, Kapra Mandal, Medchal District, Telangana 500078, India.}

\begin{abstract}
 Gravitational waves offer a key insight into the viability of classes of gravitational theories beyond general relativity. The observational constraints on their speed of propagation can provide strong constraints on generalized classes of broader gravitational frameworks. In this work, we reconsider the general class of Gauss-Bonnet theories in the context of teleparallel gravity, where the background geometry is expressed through torsion. We perform tensor perturbations on a flat FLRW background, and derive the gravitational wave propagation equation. We find that gravitational waves propagate at the speed of light in these classes of theories. We also derive the distance-duality relationship for radiation propagating in the gravitational wave and electromagnetic domains.
\end{abstract}

\maketitle

\section{Introduction} \label{SEC-I}

The standard model of cosmology, $\Lambda$CDM, has been the best way to unify astrophysical and cosmological observations over the past few decades, giving an accurate description of phenomena at these energy scales  \cite{Peebles:2002gy,Copeland:2006wr}. In this scenario, a cosmological constant ($\Lambda$) drives late-time accelerated cosmic expansion~ \cite{Riess:1998cb,Perlmutter:1998np} while gravitational interactions are explained by general relativity (GR) throughout all cosmic times whereas cold dark matter (CDM) plays  a key role in structure formation in the primordial Universe giving the seeds of large scale structure formation as well as a stabilizing agent to sustain galactic systems at later times  \cite{Baudis:2016qwx,Bertone:2004pz}. Fundamental questions on the physicality of the $\Lambda$CDM model surrounded the model since its inception such as through the fine tuning problem of the cosmological constant  \cite{Weinberg:1988cp}. Putting these features aside, $\Lambda$CDM has increasingly been challenged by recent observations  \cite{DiValentino:2025sru,DiValentino:2020vhf,DiValentino:2020zio,DiValentino:2020vvd} where estimates of its cosmological parameters increasingly appear in disharmony when early- and late-time parameter constraints are compared against each other.

The open question of tensions in $\Lambda$CDM cosmology has spurred a plethora of modified cosmological scenarios which collectively attempt to address various aspects of the open challenges of concordance cosmology. The vast array of potential new expressions of dynamical freedom at the cosmological scale have produced a battery of new classes of cosmological models  \cite{Clifton:2011jh,CANTATA:2021asi,Bahamonde:2021gfp,AlvesBatista:2021eeu,Addazi:2021xuf,Capozziello:2011et}. One interesting subclass of these models is grounded in the exchange of the foundational geometric framework of the description of gravitational interactions. One form of this is teleparallel gravity (TG) where the curvature associated with the Levi-Civita connection is substituted with the torsion expressed through the teleparallel connection so that the underlying geometric framework is transformed  \cite{Bahamonde:2021gfp,Krssak:2018ywd,Cai:2015emx}. There are a number of advantages of this framework such as having a well defined gravitational energy momentum tensor  \cite{Aldrovandi:2013wha}, as well as others which may offer a new channel to construct physical classes of cosmological models.

TG provides a unique framework on which to build curvature-free classes of gravitational models that feature a limit that is dynamically equivalent to GR, while satisfying metricity, called the teleparallel equivalent of general relativity (TEGR)  \cite{Maluf:2013gaa,aldrovandi1995introduction}. The Lagrangians of GR and TEGR differ by a boundary term $B$ which also plays an important role in certain modified gravity theories. Following the rationale of regular formulations of gravity, TEGR can easily be modified in a variety of different classes of teleparallel cosmological models. The most direct manner of modifying TEGR is through $f(T)$ gravity  \cite{Ferraro:2006jd,Ferraro:2008ey,Bengochea:2008gz,Linder:2010py,Chen:2010va,Bahamonde:2019zea,Paliathanasis:2017htk,Farrugia:2020fcu,Bahamonde:2021srr,Bahamonde:2020bbc,Bahamonde:2022ohm,Bahamonde:2020lsm} where the TEGR Lagrangian is functionally generalized analogous to the logic of $f(R)$ gravity  \cite{Sotiriou:2008rp,DeFelice:2010aj,Capozziello:2011et}. Here, the torsion scalar $T$ represents the TEGR Lagrangian. There are a variety of alternative directions in which TEGR can be generalized including the incorporation of the boundary term in $f(T,B)$ gravity  \cite{Bahamonde:2015zma,Capozziello:2018qcp,Bahamonde:2016grb,Farrugia:2018gyz,Bahamonde:2016cul,Wright:2016ayu}, possible second order scalar-tensor theories  \cite{Bahamonde:2019shr,Bahamonde_2020,Bahamonde:2020cfv,Bernardo:2021qhu,Bahamonde:2021dqn,Bernardo:2021bsg,Dialektopoulos:2021ryi,Capozziello:2023foy,Ahmedov:2023num,Ahmedov:2023lot,Ahmedov:2024aez,Capozziello:2022zzh,Capozziello:2022zzh}, as well as Gauss-Bonnet gravity generalizations through $F(T,T_G)$ gravity  \cite{Kofinas:2014daa,Kofinas:2014owa,Kofinas:2014aka,delaCruz-Dombriz:2017lvj,delaCruz-Dombriz:2018nvt,Kadam_2023,KADAM2024169563}.

Gauss-Bonnet extensions of gravity offer an interesting extension to TEGR where the Gauss-Bonnet, which is a total divergence term in $3+1$ dimensions and at linear level, is generalized in the functional form of the gravitational action. The teleparallel analog of the Gauss-Bonnet scalar $T_G$ was first identified in Ref.~ \cite{Kofinas:2014owa}, where $F(T,T_G)$ was established together with its field equations. The background cosmological equations were immediately presented in Ref.~ \cite{Kofinas:2014daa} where some specific example models were explored. The dynamical system of the background cosmological setting was finally probed in Ref.~ \cite{Kofinas:2014aka} which already showed a strong indication of physical critical points and a stable evolution profile. The behavior of potential early time cosmological models was studied in Refs.~ \cite{delaCruz-Dombriz:2017lvj,delaCruz-Dombriz:2018nvt} where an exhaustive study of potential very early-type solutions was established.

The natural next phase of probing the cosmological structure of $F(T,T_G)$ cosmologies is to consider the perturbative section. In this first exploration, we consider the tensor perturbations of the framework. This is an important first step since it establishes the speed of gravitational waves over cosmological backgrounds, which has an immediate comparison potential with observational data. Indeed, the multimessenger observations of the gravitational wave event GW170817  \cite{LIGOScientific:2017vwq} and its electromagnetic counterpart GRB170817A  \cite{Goldstein:2017mmi} placed severe constraints on the potential deviations of the gravitational wave speed from the speed of light at one part in $10^{15}$. This has disqualified many modified cosmological models, such as a vast array of Horndeski gravity cosmologies \cite{Horndeski:1974wa,Kobayashi:2019hrl,Ezquiaga_2017,Creminelli_2017,Sakstein_2017,Baker_2017}. Building on this background, we consider the tensor perturbations of $F(T,T_G)$ cosmology in this work and establish the gravitational wave propagation equation for that setting. In Sec.~\ref{SEC-II} we start with the technical details of the theory which include the Friedmann equations of the theory. In Sec.~\ref{sec:GWPE} the tensor perturbations are considered and the gravitational wave propagation equation derived. We include an analysis of this equation by establishing the amplitude and tensor excess speed parameters. We also consider the luminosity distances of gravitational waves in this context against the electromagnetic radiation counterpart. Finally, in Sec.~\ref{conc}, we summarize the results obtained in this work, highlight the key implications of the formalism under consideration, and outline possible directions for future research.

\section{\texorpdfstring{$F(T,T_G)$}{} Gravity and Cosmology} \label{SEC-II}

In theories of gravity,  the structure of interest is space-time manifold  \cite{Misner1973}. We consider a four dimensional  $ C^{\infty}$  manifold $M$  which is connected, Hausdorff and paracompact endowed with a locally Lorentzian metric $g=\{g_{\alpha\beta}\}$ and admitting absolute parallelism. That is to say, it has four linearly independent parallel vector fields, $e =\{ e_a \}=\{e_a^{\,\,\,\alpha}\}$ at each point of space-time called vierbeins which define the teleparallel connection as

\begin{equation}
\Gamma^{\gamma}_{\,\,\,\alpha\beta}:=e_{a}^{\,\,\,\gamma}
\partial_{\beta}e^{a}_{\,\,\,\alpha}+e_{a}^{\,\,\,\gamma} ~\omega^a~_{b\beta}~e^{a}_{\,\,\,\alpha}
\label{structurefun}\,
\end{equation}
where  $e^{*} =\{ e^a \}=\{e^a_{\,\,\,\alpha}\}$ is also a set of four parallel vector fields, which is dual to $e$. Additionally, $~\omega^a~_{b\beta}$ are components of connection one-forms $~\omega^a~_{b}$ called spin connection which help defining parallel transport and contain information of inertial effects. The Latin indices $a,b,$... are called ``Lorentz indices" and Greek indices $\alpha, \beta,$... are ``space-time indices" which relate the components to the local basis and the global basis respectively.  Commutation relations for vierbeins are given by
\begin{equation}
[e_{a},e_{b}]:=C^{c}_{\,\,\,ab}e_{c}\,,
\label{strucof}
\end{equation}
where  the structure coefficient functions $C^{c}_{\,\,\,ab}$ are written as
\begin{equation}
C^{c}_{\,\,\,ab}=e_{a}^{\,\,\,\alpha}
e_{b}^{\,\,\,\beta}(e^{c}_{\,\,\,\alpha,\beta}-e^{c}_{\,\,\,\beta,\alpha})
\label{structurecoeff}\,,
\end{equation}
 
An orthogonality condition is imposed on vierbeins using the metric tensor as
$g(e_a,e_b)=\eta_{ab}$, where
$\eta_{ab}=\text{diag}(-1,1,1,1)$, and therefore we get the
relation

\begin{equation}
\label{metrdef}
g_{\alpha\beta} =\eta_{ab}\, e^a_{\,\,\,\alpha}  \, e^b_{\,\,\,\beta},
\end{equation}

and indices $a,b,...$ are raised/lowered with the Minkowski metric $\eta_{ab}$ and space-time components of any tensor are obtained by contracting Lorentz indices with vierbeins or dual vierbeins.
    
 It is crucial to recognize that, for a specific metric as indicated in Eq. (\ref{metrdef}), the vierbeins associated with that metric are not unique. To ensure that the theory remains covariant, it is necessary to choose the spin connection carefully to offset the selected vierbeins. Additionally, direct incorporation of Minkowskian geometry can lead to misleading inertial effects. In particular, as shown in Eq. (\ref{metrdef}) , if the global space-time is Minkowski, this relationship holds only when the vierbeins correspond precisely to the local Lorentz transformations. To accurately address these inertial contributions, the inertial spin connection is employed, which is flat and conforms to the relation 

\begin{equation}
\partial _{[\alpha }\omega ^{a}_{\phantom {a}|b|\beta ]} + \omega ^{a}_{\phantom {a}c[\alpha } \, \omega ^{c}_{\phantom {C}|b|\beta ]} = 0\, \label{flatspin}
\end{equation} 

where square brackets symbolize antisymmetrization. To foster consistency in flat space-times, it would be beneficial to establish the inertial spin connection as

\begin{equation}
    \omega ^{a}_{\phantom {a}b\alpha }: = \Lambda ^{a}_{\phantom {a}c}\,\partial _{\alpha }\Lambda _{b}^{\phantom {b}c}\,.\label{spindef}
\end{equation}

In this context, the notation $ \Lambda ^{a}_{\phantom {a}c}$ signifies Lorentz boosts and rotations. It is established that one can consistently select a Lorentz frame in which the spin connection is zero. Vierbeins that fulfill this condition are designated as proper tetrads  \cite{aldrovandi2004spin}, and the corresponding theory is said to be situated in the  Weitzenb\"ock gauge

The teleparallel connection (\ref{structurefun}) is not symmetric in the last two indices. Its torsion is thus given by
\begin{align}
T^{\gamma}_{\;\;\alpha\beta} 
:=\,& \Gamma^{\gamma}_{\;\;\beta\alpha} - \Gamma^{\gamma}_{\;\;\alpha\beta}
\label{torsionten}
\end{align}
One can see that the components of torsion tensor and the connection coefficients in  Weitzenb\"ock gauge are related via 
\begin{equation}
T^{a}_{\,\,\,bc}=
-C^{a}_{\,\,\,bc}\,.
\end{equation}

The contortion tensor is defined as the difference of Levi-Civita connection and teleparallel connection and, since Levi-Civita connection is metric, has the expression  \cite{Bahamonde:2021gfp}

\begin{equation}\label{contortion}
 \mathcal{K}^{\alpha \beta}_{\,\,\,\,\,\gamma}:= \frac{1}{2}(T^{\beta \alpha}_{\,\,\,\,\,\gamma}+T_{\gamma}^{~~\alpha \beta}-T^{\alpha \beta}_{\,\,\,\,\,\gamma}).   
\end{equation}

The absolute parallelism condition requires $R_{\,\,\,bcd}^{a}=0$,
 which holds in all frames and which is exactly (\ref{flatspin}). The Ricci scalar
$\bar{R}$ of Levi-Civita connection can be expressed
as

\begin{equation}
e\bar{R}=-eT+2\partial_{\alpha}(eT_{\beta}^{\,\,\,\beta\alpha})\,,
\label{ricciscalar}
\end{equation}
where the ``torsion scalar'' $T$ is defined as
\begin{eqnarray}
T&=&\frac{1}{4}T^{\alpha\beta\gamma}T_{\alpha\beta\gamma}+\frac{1}{2}T^{\alpha\beta\gamma
}
T_{\gamma\beta\alpha}-T_{\beta}^{\,\,\,\beta\alpha}T^{\gamma}_{\,\,\,\gamma\alpha},
\label{Tscalar}
\end{eqnarray}
and  $e:=\det{(e^{a}_{\,\,\,\alpha})}=\sqrt{|g|}$.

It is evident now that
the Einstein-Hilbert action
\begin{eqnarray}
S_{EH}=\frac{1}{2\kappa^2}\int_{M}d^{4}\!x\,e\,\bar{R},
\label{GenRelaction}
\end{eqnarray}
where  $\kappa^2= 8\pi G$, is equivalent to the following action up to boundary terms 
\begin{eqnarray}
S_{TEGR}^{(1)}
&\!\!=\!\!&-\frac{1}{2\kappa^{2}}\int_{M}\!\!d^{4}\!x\,
e\,T
\label{teleaction}
\end{eqnarray}
in the sense that varying (\ref{GenRelaction}) with respect to the metric
and varying (\ref{teleaction}) with respect to the vierbeins gives rise to the
same equations of motion
 \cite{Aldrovandi:2013wha}. This is why Einstein called the theory with torsion scalar T as Lagrangian to be the teleparallel equivalent of general relativity.

Following the same path, i.e., to express quantities related to Levi-Civita connection in terms of quantities related to  teleparallel connection up to boundary terms, the authors of \cite{Kofinas:2014owa} defined teleparallel equivalent of Gauss-Bonnet combination $\bar{G}=\bar{R}^{2}-4\bar{R}_{\alpha\beta}\bar{R}^{\alpha\beta}+\bar{R}_{\alpha\beta\gamma\delta}\bar{R}^{\alpha\beta\gamma\delta}
$  characterized by the new torsion based scalar $T_{G}$,
as well as the equations of motion of the modified gravity defined by the function $F(T,T_{G})$:
\begin{equation}
e\bar{G}
\!=\!eT_{G}\!+\!\text{total divergence},
\label{TGdef}
\end{equation}
where $\bar{G}$ is the Gauss-Bonnet term obtained from Levi-Civita
connection, and
\begin{eqnarray}
&&\!\!\!\!\!\!\!\!\!
T_G=(\mathcal{K}^{a_{1}}_{\,\,\,\,ea}\mathcal{K}^{ea_{2}}_{\,\,\,\,\,\,\,b}
\mathcal{K}^{a_{3}}_{\,\,\,\,fc}\mathcal{K}^{fa_{4}}_{\,\,\,\,\,\,\,d}
-2\mathcal{K}^{a_{1}\!a_{2}}_{\,\,\,\,\,\,\,\,\,\,a}\mathcal{K}^{a_{3}}_{
\,\,\,\,\,eb}\mathcal{K}^{e}_{\,\,fc}\mathcal{K}^{fa_{4}}_{\,\,\,\,\,\,\,\,d}
\nn\\
&& \ \ \ \ \,+2\mathcal{K}^{a_{1}\!a_{2}}_{\,\,\,\,\,\,\,\,\,\,a}\mathcal{K}^{a_{3}}_{
\,\,\,\,\,eb}\mathcal{K}^{ea_{4}}_{\,\,\,\,\,\,\,f}\mathcal{K}^{f}_{\,\,\,cd}
\nn\\
&& \ \ \ \ \,+2\mathcal{K}^{a_{1}\!a_{2}}_{\,\,\,\,\,\,\,\,\,\,a}\mathcal{K}^{a_{3}}_{
\,\,\,\,\,eb}\mathcal{K}^{ea_{4}}_{\,\,\,\,\,\,\,c,d})\delta^{\,a\,b\,c\,d}_{a_{1}a_{2}a_{3}a_{4}}\,,
\label{TG}
\end{eqnarray}

with the generalized delta $\delta^{\,a\,b\,c\,d}_{a_{1}a_{2}a_{3}a_{4}}$ being the determinant of the Kronecker
deltas. It is noteworthy that the Latin indices are Lorentz. Therefore, $T_G$ is considered teleparallel equivalent of $\bar{G}$, in the sense
that the action for arbitrary dimension $D$.
\begin{equation}
S_{tel}^{(2)}
=\frac{1}{2\kappa_{D}^{2}}\int_{M}\!\!d^{D}\!x\,
e\,T_{G}\,,
\label{teleaction2}
\end{equation}
when varied in terms of the D-dimensional vierbeins (called vielbeins) gives exactly the same equations with the
action
\begin{equation}
S_{GB}
=\frac{1}{2\kappa_{D}^{2}}\int_{M}\!\!d^{D}\!x\,
e\,\bar{G}\,,
\label{GBaction2}
\end{equation}
varied with respect to the D-dimensional metric.

Since $T_G$ is also a topological invariant in four dimensions as $\bar{G}$ is so, actions with linear dependence of $T_G$ are of no interest and thus the generalization of the action of the form 
\begin{equation}
S_{TGB}=\frac{1}{2\kappa_{D}^{2}}\!\int d^{D}\!x\,e\,F(T,T_G)\,,
\label{FTTGgravity}
\end{equation}
is considered, which differs from
both $F(T)$ theory as well as from $F(R,G)$ gravity
 \cite{Nojiri_2005,DEFELICE20091} . TEGR (and therefore GR) is
obtained for $F(T,T_G)=-T$, while the usual Einstein-Gauss-Bonnet theory
arises for $F(T,T_G)=-T+\alpha T_G$, with $\alpha$
the Gauss-Bonnet coupling.

Equations of motion in $D=4$ after varying the action (\ref{FTTGgravity}) with respect to vierbeins are given by  \cite{Kofinas:2014owa}
\begin{eqnarray}
&&\!\!\!\!\!\!
2(\chi ^{[ac]b}\!+\!\chi^{[ba]c}\!-\!\chi^{[cb]a})_{,c}\!+\!2(\chi^{[ac]b}\!+\!\chi^{[ba]c}\!-\!\chi^{[cb]a})C^{d}_{\,\,\,dc}\nn\\
&&\!\!\!\!\!\!\!\!+(2\chi^{[ac]d}\!+\!\chi^{dca})C^{b}_{\,\,\,cd}
\!+\!4\chi^{[db]c}C_{(dc)}^{\,\,\,\,\,\,\,\,a}\!+\!(T^{a}_{\,\,\,cd}\!+\!2\omega^{a}_{\,\,\,[cd]})\chi^{cdb}\nn\\
&&\,\,\,\,\,\,\,\,\,\,\,\,\,\,\,\,\,\,\,\,\,\,\,-(-1)^{D}h^{ab}\!+\!(F\!-\!TF_{T}\!-\!T_{G}F_{T_{G}})\eta^{ab}=0\,,
\label{gkw}
\end{eqnarray}

where
\begin{eqnarray}
&& \!\!\!\!\!\!\!
\chi^{abc}=F_{T}(\eta^{ac}\mathcal{K}^{bd}_{\,\,\,\,\,\,d}-\mathcal{K}^{bca})+F_{T_{G}}\big[\nn\\
&&\!\!\!\!\!\!\!\epsilon^{cprt}\!\big(\!2\epsilon^{a}_{\,\,\,dkf}\mathcal{K}^{bk}_{\,\,\,\,\,p}
\mathcal{K}^{d}_{\,\,\,qr}
\!\!+\!\epsilon_{qdkf}\mathcal{K}^{ak}_{\,\,\,\,\,p}\mathcal{K}^{bd}_{\,\,\,\,\,\,r}\!\!+\!
\epsilon^{ab}_{\,\,\,\,\,\,kf}\mathcal{K}^{k}_{\,\,\,dp}\mathcal{K}^{d}_{\,\,\,qr}\!\big)\!
\mathcal{K}^{qf}_{\,\,\,\,\,\,t}\nn\\
&&\!\!\!\!\!\!\!+\epsilon^{cprt}\epsilon^{ab}_{\,\,\,\,\,\,kd}\mathcal{K}^{fd}_{\,\,\,\,\,\,p}
\big(\mathcal{K}^{k}_{\,\,fr,t}\!-\!\frac{1}{2}\mathcal{K}^{k}_{\,\,fq}C^{q}_{\,\,\,tr}
\!\!+\!\omega^{k}_{\,\,\,qt}\mathcal{K}^{q}_{\,\,fr}\!\!+\!\omega^{q}_{\,\,fr}\mathcal{K}^{k}_{\,\,qt}\big)\nn\\
&&\!\!\!\!\!\!\!+\epsilon^{cprt}\epsilon^{ak}_{\,\,\,\,\,\,df}\mathcal{K}^{df}_{\,\,\,\,p}
\big(\mathcal{K}^{b}_{\,\,kr,t}\!-\!\frac{1}{2}\mathcal{K}^{b}_{\,\,kq}C^{q}_{\,\,\,tr}
\!\!+\!\omega^{b}_{\,\,\,qt}\mathcal{K}^{q}_{\,\,kr}\!\!+\!\omega^{q}_{\,\,\,kr}\mathcal{K}^{b}_{\,\,qt}\big)\big]
\nn\\
&&\!\!\!\!\!\!\!+F_{T_{G}}\epsilon^{cprt}\epsilon^{a}_{\,\,\,kdf}\Big[
\frac{1}{F_{T_{G}}}\big(F_{T_{G}}\mathcal{K}^{bk}_{\,\,\,\,\,p}
\mathcal{K}^{df}_{\,\,\,\,\,r}\big)_{,t}\!+\!
C^{q}_{\,\,\,pt}\mathcal{K}^{bk}_{\,\,\,\,\,[q}\mathcal{K}^{df}_{\,\,\,\,\,r]}
\nn\\
&&\!\!\!\!\!\!\!
+(\omega^{b}_{\,\,\,qp}\mathcal{K}^{qk}_{\,\,\,\,\,\,r}\!+\!\omega^{k}_{\,\,\,qp}\mathcal{K}^{bq}_{\,\,\,\,\,\,r})
\mathcal{K}^{df}_{\,\,\,\,\,t}
\!+\!(\omega^{d}_{\,\,\,qp}\mathcal{K}^{qf}_{\,\,\,\,\,\,t}\!+\!\omega^{f}_{\,\,\,qp}\mathcal{K}^{dq}_{\,\,\,\,\,\,t})
\mathcal{K}^{bk}_{\,\,\,\,\,r}\Big]\label{kas}\nn\\
\end{eqnarray}

and
\begin{equation}
h^{ab}=F_{T}\epsilon^{a}_{\,\,\,kcd}\epsilon^{bpqd}\mathcal{K}^{k}_{\,\,\,fp}
\mathcal{K}^{fc}_{\,\,\,\,\,\,q}\,,
\label{hab22}
\end{equation}
and $C^{c}_{\,\,\,ab}$ are structure coefficients defined in (\ref{strucof}). Furthermore,
the denotations $F_{T}=\partial F/\partial T$, $F_{T_{G}}=\partial F/\partial T_{G}$ are used and 
the (anti)symmetrization symbol contains the factor $\frac{1}{2}$. In Weitzenb\"ock gauge , that is, in the frame where $\omega^{a}_{\,\,\,bc}=0$, Eq. (\ref{kas}) reduces to 
\begin{eqnarray}
&& \!\!\!\!\!\!\!
\chi^{abc}=F_{T}(\eta^{ac}\mathcal{K}^{bd}_{\,\,\,\,\,\,d}-\mathcal{K}^{bca})+F_{T_{G}}\big[\nn\\
&&\!\!\!\!\!\!\!\epsilon^{cprt}\!\big(\!2\epsilon^{a}_{\,\,\,dkf}\mathcal{K}^{bk}_{\,\,\,\,\,p}
\mathcal{K}^{d}_{\,\,\,qr}
\!\!+\!\epsilon_{qdkf}\mathcal{K}^{ak}_{\,\,\,\,\,p}\mathcal{K}^{bd}_{\,\,\,\,\,\,r}\!\!+\!
\epsilon^{ab}_{\,\,\,\,\,\,kf}\mathcal{K}^{k}_{\,\,\,dp}\mathcal{K}^{d}_{\,\,\,qr}\!\big)\!
\mathcal{K}^{qf}_{\,\,\,\,\,\,t}\nn\\
&&\,\,\,\,\,\,\,\,\,\,\,\,\,\,\,\,\,\,\,\,\,\,\,\,\,\,\,\,\,\,\,\,\,\,\,\,\,\,\,\,\,\,\,\,
+\epsilon^{cprt}\epsilon^{ab}_{\,\,\,\,\,\,kd}\mathcal{K}^{fd}_{\,\,\,\,\,\,p}
\big(\mathcal{K}^{k}_{\,\,fr,t}\!-\!\frac{1}{2}\mathcal{K}^{k}_{\,\,fq}C^{q}_{\,\,\,tr}\big)\nn\\
&&\,\,\,\,\,\,\,\,\,\,\,\,\,\,\,\,\,\,\,\,\,\,\,\,\,\,\,\,\,\,\,\,\,\,\,\,\,\,\,\,\,\,\,
+\epsilon^{cprt}\epsilon^{ak}_{\,\,\,\,\,\,df}\mathcal{K}^{df}_{\,\,\,\,p}
\big(\mathcal{K}^{b}_{\,\,kr,t}\!-\!\frac{1}{2}\mathcal{K}^{b}_{\,\,kq}C^{q}_{\,\,\,tr}\big)\big]
\nn\\
&&\!\!+\epsilon^{cprt}\epsilon^{a}_{\,\,\,kdf}\Big[
\big(F_{T_{G}}\mathcal{K}^{bk}_{\,\,\,\,\,p}
\mathcal{K}^{df}_{\,\,\,\,\,r}\big)_{,t}\!+\!
F_{T_{G}}C^{q}_{\,\,\,pt}\mathcal{K}^{bk}_{\,\,\,\,\,[q}\mathcal{K}^{df}_{\,\,\,\,\,r]}\Big]\label{ksi}, 
\end{eqnarray} 
Eq. (\ref{hab22}) is unchanged and the equation of motion (\ref{gkw})
is reduced to 
\begin{eqnarray}
&&\!\!\!\!\!\!
2(\chi^{[ac]b}\!+\!\chi^{[ba]c}\!-\!\chi^{[cb]a})_{,c}\!+\!2(\chi^{[ac]b}\!+\!\chi^{[ba]c}\!-\!\chi^{[cb]a})C^{d}_{\,\,\,dc}\nn\\
&&\!\!\!\!\!\!\!\!+(2\chi^{[ac]d}\!+\!\chi^{dca})C^{b}_{\,\,\,cd}
\!+\!4\chi^{[db]c}C_{(dc)}^{\,\,\,\,\,\,\,\,a}\!+\!T^{a}_{\,\,\,cd}\chi^{cdb}-h^{ab}\nn\\
&&\,\,\,\,\,\,\,\,\,\,\,\,\,\,\,\,\,\,\,\,\,\,\,\,\,\,\,\,\,\,\,\,\,\,\,\,\,\,\,\,\,\,\,\,\,\,\,\,\,\,
+(F\!-\!TF_{T}\!-\!T_{G}F_{T_{G}})\eta^{ab}=0\,,
\label{kal}
\end{eqnarray}

\label{Fcosmo}

For cosmological applications, we proceed with the total action
\begin{eqnarray}
S_{total} =\frac{1}{2\kappa^{2}}\!\int d^{4}\!x\,e\,F(T,T_G)\,+S_{M}\,,
\label{fGBtelaction}
\end{eqnarray}
where $S_{M}$ is the matter Lagrangian which after variation yields a matter energy-momentum tensor
$\Theta^{\alpha\beta}$.
We assume a spatially flat Friedmann--Lemaître--Robertson--Walker cosmological ansatz,
\begin{equation}
ds^{2}=-N^2
dt^{2}+a^{2}\delta_{\hat{i}\hat{j}}dx^{\hat{i}}dx^{\hat{j}}\,,
\label{metriccosmo}
\end{equation}
where $a$ is the scale factor and $N$ is the lapse function (the
hat indices run in the three spatial coordinates).
This metric arises from the diagonal vierbeins
\begin{equation}
\label{vierbeincosmo}
e^{a}_{\,\,\,\alpha}=\text{diag}(N,a,a,a)   
\end{equation}
through (\ref{metrdef}). We emphasize that this ansatz is compatible with the Weitzenb\"ock gauge  \cite{Bahamonde:2021gfp}. 
Considering as
usual $N=1$ and inserting the ansatz (\ref{vierbeincosmo}) into (\ref{Tscalar})  and (\ref{TG}), we obtain 
\begin{align}
T &= 6\frac{\dot{a}^{2}}{a^{2}} = 6H^2, \label{Tcosmo} \\
T_G &= 24\frac{\dot{a}^2}{a^2} \cdot \frac{\ddot{a}}{a} 
= 24H^2\big(\dot{H} + H^2\big). \label{TGcosmo}
\end{align}

where $H=\frac{\dot{a}}{a}$ is the Hubble parameter and dots denote
differentiation with respect to $t$. Additionally, inserting
(\ref{vierbeincosmo}) into the general equations of motion
(\ref{kal}), we get the Friedmann
equations \cite{Kofinas:2014daa}
\begin{equation}
F-12H^{2}F_{T}-T_{G}F_{T_{G}}+24H^{3}\dot{F_{T_{G}}}=2\kappa^{2}\rho
\label{eqmN}
\end{equation}
\begin{align}
F &- 4(\dot{H} + 3H^2)F_T - 4H\dot{F}_T 
- T_G F_{T_G} + \frac{2}{3H}T_G \dot{F}_{T_G} \nonumber\\
&+ 8H^2 \ddot{F}_{T_G} = -2\kappa^2 p\,
\label{eqma}
\end{align}

where we have considered $\Theta^{\alpha\beta}$ to be the energy momentum tensor of  a perfect fluid with energy
density $\rho$ and pressure $p$ (i.e. $\Theta^{00}=\rho$,
$\Theta^{\hat{i}\hat{j}}=\frac{p}
{a^{2}}\delta^{\hat{i}\hat{j}}$, $\Theta^{\hat{i}}_{\,\,\hat{i}}=3p$). In the above expressions
notations $\dot{F_{T}}=F_{TT}\dot{T}+F_{TT_{G}}\dot{T}_{G}$,
$\dot{F_{T_{G}}}=F_{TT_{G}}\dot{T}+F_{T_{G}T_{G}}\dot{T}_{G}$ and 
$\ddot{F_{T_{G}}}=F_{TTT_{G}}\dot{T}^{2}+2F_{TT_{G}T_{G}}\dot{T}
\dot{T}_{G}+F_{T_{G}T_{G}T_{G}}\dot{T}_{G}^{\,\,2}+
F_{TT_{G}}\ddot{T}+F_{T_{G}T_{G}}\ddot{T}_{G}$ are used,
with $F_{TT}$, $F_{TT_{G}}$,\,... denoting multiple partial differentiations
of $F_T$ with respect to $T$, $T_{G}$.
Finally, $\dot{T}$, $\ddot{T}$ and $\dot{T}_{G}$, $\ddot{T}_{G}$ are obtained
by differentiating (\ref{Tcosmo}) and (\ref{TGcosmo}), respectively, with
respect to time.

\section{Gravitational wave propagation equation (GWPE)}\label{sec:GWPE}
In dynamical theories of gravity, tensor perturbations correspond to gravitational waves (GWs) that, when field equations are expanded into orders, yield propagation equations for gravitational waves. This equation can be used to probe consistent modifications and can be used for comparison with GR.  In  \cite{Farrugia:2018gyz} it was shown that GWs have only two polarization modes on a Minkowski background. In what follows, we shall derive the cosmological GWPE and analyze the implications.

In GR and its curvature based extensions the GWPE is obtained by taking tensor perturbation about background cosmological metric tensor as $g_{\alpha\beta} \xrightarrow{pert.}g_{\alpha\beta}+\delta g_{\alpha\beta}$ where $g_{\alpha\beta}$ is background cosmological ansatz and $\delta g_{\alpha\beta}$ in GR carries two degrees of freedom of GWs.
Thanks to relation (\ref{metrdef}), the perturbative approaches of metric based theories can be directly transferred to vierbein based ones.  Perturbations of the form $e^a_{\,\,\,\alpha}\xrightarrow{pert.}e^a_{\,\,\,\alpha}+\delta e^a_{\,\,\,\alpha}$ are taken where $|\delta e^a_{\,\,\,\alpha}|\ll1 $. Although general perturbations are not free from gauge problems, tensor perturbations are gauge invariant in both metric and vierbein infinitesimal diffeomorphisms at first order \cite{Bardeen1980,Chen:2010va,heisenberg2023}.
At first order tensor perturbations in metric are of the form $\delta g_{\alpha\beta}=a^2 \delta^{\hat{i}}_{\,\alpha} \delta^{\hat{j}}_{\,\beta} h_{\hat{i}\hat{j}}$,  where $\hat{i} , \hat{j}$ are spatial. We need to consider perturbations in vierbeins to reproduce this metric at first order. It is trivial to see that this can be done with the vierbein choice
\begin{equation}
    \delta e^{\hat{k}}_{\,\,\alpha}=\frac{a}{2} \delta^{\hat{i}}_{\,\alpha}\delta^{\hat{k}\hat{j}} h_{\hat{i}\hat{j}}\label{tentetrad} 
\end{equation}

where all of $\hat{i} , \hat{j}, \hat{k}$ are spatial and $h_{\hat{i}\hat{j}}$ satisfy the gauge conditions $ \text{traceless:} \quad  h^{\hat{i}}_{\ \hat{i}} = 0$ and $\text{transverse:} \quad  \partial^{\hat{i}} h_{\hat{i}\hat{j}} = 0$. It is crucial to mention this perturbation choice remains in Weitzenb\"ock gauge, even at perturbative order  \cite{Bahamonde:2020lsm}

A general GWPE in most modified gravity theories usually takes the form  \cite{Saltas_2014}:
\begin{equation}\label{G_GWPE}
\ddot{h}_{\hat{i}\hat{j}}+\left(3+\alpha_{M}\right)H\dot{h}_{\hat{i}\hat{j}}+\left(1+\alpha_{T}\right)\frac{k^{2}}{a^{2}}h_{\hat{i}\hat{j}}=0\,,
\end{equation}
where dots denote differentiation with respect to cosmic time, $\displaystyle H=\frac{\dot{a}}{a}$ is the Hubble parameter, $\displaystyle \alpha_{M}=\frac{1}{HM_{\ast}^{2}}\frac{dM_{\ast}^{2}}{dt}$ is Planck mass running rate, and $\alpha_{T}=c_{T}^{2} - 1$ is the tensor excess speed. The GWPE in Eq.~(\ref{G_GWPE}) is being considered in its Fourier domain, along with a source-free scenario.

Now to find GWPE there are two paths one can take. First would be to linearize the field equations around the background. This generally works for any gravity theory and one can go to higher order perturbations too but this can get algebraically heavy and it is harder to isolate physical degrees of freedom. On the other hand, one can perturb the action of the theory instead of equations of motion. Then, the perturbation equations are derived by varying the perturbed action with respect to the perturbed field. This provides a framework to identify physical degrees of freedom and naturally avoids constraints that do not propagate. We will choose the latter path.
 Inserting perturbed relations \eqref{appen} in (\ref{Tscalar}) and (\ref{TG}) and expanding the action (\ref{fGBtelaction}), we get the second order action after integration by parts and using gauge conditions to be 
\begin{equation} 
    \mathcal{S}^{(2)}_{\rm T}=\frac{1}{2\kappa^{2}}\! \int d t d ^3 x \, \,\frac{a^3}{2}\mathcal{C}_{tensor} \left[\, \dot{h}_{\hat{i}\hat{j}}^2 - a^{-2}(\bm{\nabla} h_{\hat{i}\hat{j}})^2\,\right]\,
    \label{eq:ten_pert_action}
\end{equation}
where
$\mathcal{C}_{tensor}=-F_T+4H\dot{F}_{T_G}$. The overdot represents the derivative with respect to cosmic time t. We have considered the matter content to be a perfect fluid with no anisotropic stress, therefore it does not source tensor perturbations. We thus set the matter Lagrangian to zero in the tensor sector.
 
Varying the action (\ref{eq:ten_pert_action}) with respect to $h_{\hat{i}\hat{j}}$ yields the GWPE

\begin{equation}\label{TTG_GWPE}
\ddot{h}_{\hat{i}\hat{j}}+\left(3H+\frac{-\dot{F}_T+4\dot{H}\dot{F}_{T_G}+4H\ddot{F}_{T_G}}{-F_T+4H\dot{F}_{T_G}}\right)\dot{h}_{\hat{i}\hat{j}}+\frac{k^{2}}{a^{2}}h_{\hat{i}\hat{j}}=0\,,
\end{equation}

where we have made sure that on shell conditions are satisfied. Comparing Eq. (\ref{TTG_GWPE}) with (\ref{G_GWPE}), we have the effective mass 
$M_{\ast}^{2}=\kappa^{-2}(-F_T+4H\dot{F}_{T_G})\,$  and the excess tensor speed $\,
\alpha_T=0$. Thus, gravitational waves propagate with the speed of light. Additionally, to avoid ghostlike behavior of perturbations, the condition $\mathcal{C}_{tensor}>0$ must be satisfied.

The modification parameters outlined in Eq. (\ref{G_GWPE}) play an important role in the detail of the waveform, altering both its amplitude and phase through the $\alpha_M$ and $\alpha_T$, respectively. By expanding the waveform around its general relativity limit, we can gain valuable insights into these effects. This is visualized via the relation  \cite{Bahamonde_2020}  

\begin{equation}
h_{\text{TGB}}\sim h_{\rm GR}\;\underset{{\scriptscriptstyle \rm Amplitude}}{\underbrace{e^{-\frac{1}{2}\int\alpha_{M}\mathcal{H}d\eta}}}\;
\end{equation}
where $\eta=\int dt/a$ denotes conformal time, $\mathcal{H}=a'/a$ is the conformal Hubble parameter and primes represent derivatives with respect to conformal time. Modification of GWPE results in change of GW luminosity distance  \cite{Belgacem_2019}. In these cases, the luminosity distance for the GWs is related to their electromagnetic counterpart by  \cite{Ezquiaga_2017}
\begin{equation}
\frac{d_L^{g}(z)}{d_L^{EM}(z)} = \exp\left[\frac{1}{2}\int_0^z \frac{\alpha_M}{1+z'}\,dz'\right]\,,
\end{equation}
from which the damping of GWs against $z$ can be used to constrain the frictional term $\alpha_M$. This can be done using standard sirens which next generation gravitational wave detectors are going to have as main motive.

The gravitational waves events viz. GW170817 and its electromagnetic counterpart GRB170817A set limits for values of $\alpha_T$. For flat cosmological models this parameter is effectively zero. Therefore, the theory is consistent while satisfying observational constraints.

\section{Conclusion}\label{conc}
In this work, tensor perturbations around an $F(T,T_G)$ class of modified cosmological models is considered. We build on previous work on the background evolution for these cosmologies, and develop the GWPE including the parametrization of both the speed of propagation and the amplitude term. We find that GW propagation occurs at the speed of light, while the amplitude term suffers a modification. The speed constraint aligns with current observations, while amplitude modifications remain an open question.

The ongoing advancement of gravitational-wave multimessenger astronomy opens a novel observational avenue for probing the fundamental aspects of nature, particularly in the contexts of cosmology and gravitational physics. In this work we focused on the $F(T,T_G)$ background and analyzed tensor perturbations thereof leading to a second order propagation equation of gravitational waves. A fundamental aspect of this investigation is that the gravitational theory in question dictates both the intrinsic characteristics of gravitational waves such as their propagation speed and polarization modes and the properties of the cosmological background through which they travel, notably the expanding universe.  Consequently, observations of gravitational wave features can be employed to constrain and differentiate among various gravitational theories.  For example in the context of $f(T)$ gravity, the propagation equation was derived in  \cite{Cai_2018} and luminal speed as well as two tensor modes were confirmed. In the case of $F(T,T_G)$ gravity, we showed that the gravitational waves propagate with the speed of light which is in accordance with observational constraints put by GW170817 \cite{LIGOScientific:2017vwq}
 and its EM counterpart  GRB170817A  \cite{Goldstein:2017mmi}. We emphasize that the modified teleparallel Gauss-Bonnet gravity is a novel extension compared to a similar extension of the Gauss-Bonnet invariant since the gravitational wave propagation speed deviates from the speed of light  \cite{DeFelice:2010sh}, therefore constraining various models in the light of limit posed by observations. This study is the first of many ways to constrain various models pertaining to $F(T,T_G)$ exploiting the perturbative regime, one being  $\mathcal{C}_{tensor}>0$ that follows from the stability condition avoiding ghosts in the action (\ref{eq:ten_pert_action}). One can follow the recipe outlined in  \cite{Ezquiaga_2017, Nishizawa_2018} to constrain various functional forms of $F(T,T_G)$.   
\section*{Acknowledgments}B.M. acknowledges the support of Anusandhan National Research Foundation(ANRF), Science \& Engineering Research Board(SERB), DST for the grant (File No: CRG/2023/000475). This article is also based upon work from COST Action CA21136 Addressing observational tensions in cosmology with systematics and fundamental physics (CosmoVerse) supported by COST (European Cooperation in Science and Technology).
\appendix
\section{} \label{appen}

We have quantities perturbed up to second order as:

\begin{eqnarray*}
&&e^0_{\ \alpha}dx^\alpha= dt,\\ 
&&e^a_{\ \alpha} dx^\alpha =a~\delta_{a\hat{i}}
\left(\delta_{\hat{i}\hat{j}}+\frac{1}{2}h_{\hat{i}\hat{j}} \right) dx^{\hat{j}},\\
&&e_0^{\ \alpha} \frac{\partial}{\partial x^\alpha}=\frac{\partial}{\partial t},\\
&&e_a^{\ \alpha}\frac{\partial}{\partial x^\alpha}= 
a^{-1} \delta _{a\hat{j}} (\delta_{\hat{i}\hat{j}}-\frac{1}{2}h_{\hat{i}\hat{j}}+\frac{1}{4}h_{\hat{i}\hat{k}}h_{\hat{k}\hat{j}})
\frac{\partial}{\partial x^{\hat{i}}},\\
&&g_{\alpha\beta}dx^\alpha dx^\beta= - dt^2 +a^2 (\delta_{\hat{i}\hat{j}} +h_{\hat{i}\hat{j}} +\frac{1}{4}h_{\hat{i}\hat{k}}h_{\hat{k}\hat{j}})
dx^{\hat{i}} dx^{\hat{j}},\\ \vspace{10cm}\\
&&g^{\alpha\beta}\frac{\partial}{\partial x^\alpha}\frac{\partial}{\partial x^\beta}=
-\left(\frac{\partial}{\partial t}\right)^2
+\frac{1}{a^{2}}(\delta_{\hat{i}\hat{j}} -h_{\hat{i}\hat{j}} +\frac{3}{4}h_{\hat{i}\hat{k}}h_{\hat{k}\hat{j}})
\frac{\partial}{\partial x^{\hat{i}}}\frac{\partial}{\partial x^{\hat{j}}}.
\end{eqnarray*}

We have
\begin{eqnarray*}
&&e = a^3 \left(1-\frac{1}{4}h_{\hat{i}\hat{j}}^2\right).
\end{eqnarray*}
The space-time components of the torsion tensor are:
\begin{eqnarray*}
&&T^0_{\ 0\hat{i}}= 0, \\
&&T^0_{\ \hat{i}\hat{j}}= 0, \\
&&T^{\hat{i}}_{\ 0\hat{j}} = H \delta_{\hat{i}\hat{j}}+ \frac{1}{2}\dot h_{\hat{i}\hat{j}} -\frac{1}{4}h_{\hat{i}\hat{k}} \dot h_{\hat{k}\hat{j}}, \\
&&T^{\hat{i}}_{\ \hat{j}\hat{k}} =\frac{1}{2}\partial_{\hat{j}} h_{\hat{i}\hat{k}}- \frac{1}{2}\partial_{\hat{k}} h_{\hat{i}\hat{j}}-\frac{1}{4}h_{\hat{i}\hat{l}}\partial_{\hat{j}} h_{\hat{l}\hat{k}}
+\frac{1}{4}h_{\hat{i}\hat{l}} \partial_{\hat{k}} h_{\hat{l}\hat{j}},\\
&&T^\alpha_{\ 0 \alpha} = 3H -\frac{1}{4}h_{\hat{i}\hat{j}} \dot h_{\hat{i}\hat{j}},\\
&&T^\alpha_{\ \hat{i}\alpha}= -\frac{1}{4}h_{\hat{j}\hat{k}} \partial_{\hat{i}} h_{\hat{j}\hat{k}}+\frac{1}{4} h_{\hat{j}\hat{k}} \partial_{\hat{j}} h_{\hat{k}\hat{i}}.
\end{eqnarray*}
where Einstein's summation convention is used.  

\bibliographystyle{utphys}

\bibliography{references}

\providecommand{\href}[2]{#2}\begingroup\raggedright\begin{thebibliography}{10}

\bibitem{Peebles:2002gy}
P.~J.~E. Peebles and B.~Ratra, ``{The Cosmological Constant and Dark Energy},'' \href{https://doi.org/10.1103/RevModPhys.75.559}{{\em Rev. Mod. Phys.} {\bf 75} (2003)  559--606}, \href{http://arxiv.org/abs/astro-ph/0207347}{{\tt arXiv:astro-ph/0207347}}.

\bibitem{Copeland:2006wr}
E.~J. Copeland, M.~Sami, and S.~Tsujikawa, ``{Dynamics of dark energy},'' \href{https://doi.org/10.1142/S021827180600942X}{{\em Int. J. Mod. Phys. D} {\bf 15} (2006)  1753--1936}, \href{http://arxiv.org/abs/hep-th/0603057}{{\tt arXiv:hep-th/0603057}}.

\bibitem{Riess:1998cb}
{\bf Supernova Search Team} Collaboration, A.~G. Riess {\em et al.}, ``{Observational evidence from supernovae for an accelerating universe and a cosmological constant},'' \href{https://doi.org/10.1086/300499}{{\em Astron. J.} {\bf 116} (1998)  1009--1038}, \href{http://arxiv.org/abs/astro-ph/9805201}{{\tt arXiv:astro-ph/9805201}}.

\bibitem{Perlmutter:1998np}
{\bf Supernova Cosmology Project} Collaboration, S.~Perlmutter {\em et al.}, ``{Measurements of $\Omega$ and $\Lambda$ from 42 High Redshift Supernovae},'' \href{https://doi.org/10.1086/307221}{{\em Astrophys. J.} {\bf 517} (1999)  565--586}, \href{http://arxiv.org/abs/astro-ph/9812133}{{\tt arXiv:astro-ph/9812133}}.

\bibitem{Baudis:2016qwx}
L.~Baudis, ``{Dark matter detection},'' \href{https://doi.org/10.1088/0954-3899/43/4/044001}{{\em J. Phys. G} {\bf 43} (2016) no.~4, 044001}.

\bibitem{Bertone:2004pz}
G.~Bertone, D.~Hooper, and J.~Silk, ``{Particle dark matter: Evidence, candidates and constraints},'' \href{https://doi.org/10.1016/j.physrep.2004.08.031}{{\em Phys. Rept.} {\bf 405} (2005)  279--390}, \href{http://arxiv.org/abs/hep-ph/0404175}{{\tt arXiv:hep-ph/0404175}}.

\bibitem{Weinberg:1988cp}
S.~Weinberg, ``{The Cosmological Constant Problem},'' \href{https://doi.org/10.1103/RevModPhys.61.1}{{\em Rev. Mod. Phys.} {\bf 61} (1989)  1--23}.

\bibitem{DiValentino:2025sru}
E.~Di~Valentino {\em et al.}, ``{The CosmoVerse White Paper: Addressing observational tensions in cosmology with systematics and fundamental physics},'' \href{http://arxiv.org/abs/2504.01669}{{\tt arXiv:2504.01669 [astro-ph.CO]}}.

\bibitem{DiValentino:2020vhf}
E.~Di~Valentino {\em et al.}, ``{Snowmass2021 - Letter of interest cosmology intertwined I: Perspectives for the next decade},'' \href{https://doi.org/10.1016/j.astropartphys.2021.102606}{{\em Astropart. Phys.} {\bf 131} (2021)  102606}, \href{http://arxiv.org/abs/2008.11283}{{\tt arXiv:2008.11283 [astro-ph.CO]}}.

\bibitem{DiValentino:2020zio}
E.~Di~Valentino {\em et al.}, ``{Snowmass2021 - Letter of interest cosmology intertwined II: The hubble constant tension},'' \href{https://doi.org/10.1016/j.astropartphys.2021.102605}{{\em Astropart. Phys.} {\bf 131} (2021)  102605}, \href{http://arxiv.org/abs/2008.11284}{{\tt arXiv:2008.11284 [astro-ph.CO]}}.

\bibitem{DiValentino:2020vvd}
E.~Di~Valentino {\em et al.}, ``{Cosmology Intertwined III: $f \sigma_8$ and $S_8$},'' \href{https://doi.org/10.1016/j.astropartphys.2021.102604}{{\em Astropart. Phys.} {\bf 131} (2021)  102604}, \href{http://arxiv.org/abs/2008.11285}{{\tt arXiv:2008.11285 [astro-ph.CO]}}.

\bibitem{Clifton:2011jh}
T.~Clifton, P.~G. Ferreira, A.~Padilla, and C.~Skordis, ``{Modified Gravity and Cosmology},'' \href{https://doi.org/10.1016/j.physrep.2012.01.001}{{\em Phys. Rept.} {\bf 513} (2012)  1--189}, \href{http://arxiv.org/abs/1106.2476}{{\tt arXiv:1106.2476 [astro-ph.CO]}}.

\bibitem{CANTATA:2021asi}
{\bf CANTATA} Collaboration, Y.~Akrami {\em et al.}, \href{https://doi.org/10.1007/978-3-030-83715-0}{{\em {Modified Gravity and Cosmology. An Update by the CANTATA Network}}}.
\newblock Springer, 2021.
\newblock \href{http://arxiv.org/abs/2105.12582}{{\tt arXiv:2105.12582 [gr-qc]}}.

\bibitem{Bahamonde:2021gfp}
S.~Bahamonde, K.~F. Dialektopoulos, C.~Escamilla-Rivera, G.~Farrugia, V.~Gakis, M.~Hendry, M.~Hohmann, J.~Levi~Said, J.~Mifsud, and E.~Di~Valentino, ``{Teleparallel gravity: from theory to cosmology},'' \href{https://doi.org/10.1088/1361-6633/ac9cef}{{\em Rept. Prog. Phys.} {\bf 86} (2023) no.~2, 026901}, \href{http://arxiv.org/abs/2106.13793}{{\tt arXiv:2106.13793 [gr-qc]}}.

\bibitem{AlvesBatista:2021eeu}
R.~Alves~Batista {\em et al.}, ``{EuCAPT White Paper: Opportunities and Challenges for Theoretical Astroparticle Physics in the Next Decade},'' \href{http://arxiv.org/abs/2110.10074}{{\tt arXiv:2110.10074 [astro-ph.HE]}}.

\bibitem{Addazi:2021xuf}
A.~Addazi {\em et al.}, ``{Quantum gravity phenomenology at the dawn of the multi-messenger era\textemdash{}A review},'' \href{https://doi.org/10.1016/j.ppnp.2022.103948}{{\em Prog. Part. Nucl. Phys.} {\bf 125} (2022)  103948}, \href{http://arxiv.org/abs/2111.05659}{{\tt arXiv:2111.05659 [hep-ph]}}.

\bibitem{Capozziello:2011et}
S.~Capozziello and M.~De~Laurentis, ``{Extended Theories of Gravity},'' \href{https://doi.org/10.1016/j.physrep.2011.09.003}{{\em Phys. Rept.} {\bf 509} (2011)  167--321}, \href{http://arxiv.org/abs/1108.6266}{{\tt arXiv:1108.6266 [gr-qc]}}.

\bibitem{Krssak:2018ywd}
M.~Krssak, R.~J. van~den Hoogen, J.~G. Pereira, C.~G. B\"ohmer, and A.~A. Coley, ``{Teleparallel theories of gravity: illuminating a fully invariant approach},'' \href{https://doi.org/10.1088/1361-6382/ab2e1f}{{\em Class. Quant. Grav.} {\bf 36} (2019) no.~18, 183001}, \href{http://arxiv.org/abs/1810.12932}{{\tt arXiv:1810.12932 [gr-qc]}}.

\bibitem{Cai:2015emx}
Y.-F. Cai, S.~Capozziello, M.~De~Laurentis, and E.~N. Saridakis, ``{f(T) teleparallel gravity and cosmology},'' \href{https://doi.org/10.1088/0034-4885/79/10/106901}{{\em Rept. Prog. Phys.} {\bf 79} (2016) no.~10, 106901}, \href{http://arxiv.org/abs/1511.07586}{{\tt arXiv:1511.07586 [gr-qc]}}.

\bibitem{Aldrovandi:2013wha}
R.~Aldrovandi and J.~G. Pereira, \href{https://doi.org/10.1007/978-94-007-5143-9}{{\em Teleparallel Gravity: An Introduction}}.
\newblock Springer, 2013.

\bibitem{Maluf:2013gaa}
J.~W. Maluf, ``{The teleparallel equivalent of general relativity},'' \href{https://doi.org/10.1002/andp.201200272}{{\em Annalen Phys.} {\bf 525} (2013)  339--357}, \href{http://arxiv.org/abs/1303.3897}{{\tt arXiv:1303.3897 [gr-qc]}}.

\bibitem{aldrovandi1995introduction}
R.~Aldrovandi and J.~Pereira, \href{https://doi.org/10.1142/10202}{{\em An Introduction to Geometrical Physics}}.
\newblock World Scientific, 1995.

\bibitem{Ferraro:2006jd}
R.~Ferraro and F.~Fiorini, ``{Modified teleparallel gravity: Inflation without inflaton},'' \href{https://doi.org/10.1103/PhysRevD.75.084031}{{\em Phys. Rev. D} {\bf 75} (2007)  084031}, \href{http://arxiv.org/abs/gr-qc/0610067}{{\tt arXiv:gr-qc/0610067}}.

\bibitem{Ferraro:2008ey}
R.~Ferraro and F.~Fiorini, ``{On Born-Infeld Gravity in Weitzenbock spacetime},'' \href{https://doi.org/10.1103/PhysRevD.78.124019}{{\em Phys. Rev. D} {\bf 78} (2008)  124019}, \href{http://arxiv.org/abs/0812.1981}{{\tt arXiv:0812.1981 [gr-qc]}}.

\bibitem{Bengochea:2008gz}
G.~R. Bengochea and R.~Ferraro, ``{Dark torsion as the cosmic speed-up},'' \href{https://doi.org/10.1103/PhysRevD.79.124019}{{\em Phys. Rev. D} {\bf 79} (2009)  124019}, \href{http://arxiv.org/abs/0812.1205}{{\tt arXiv:0812.1205 [astro-ph]}}.

\bibitem{Linder:2010py}
E.~V. Linder, ``{Einstein's Other Gravity and the Acceleration of the Universe},'' \href{https://doi.org/10.1103/PhysRevD.81.127301}{{\em Phys. Rev. D} {\bf 81} (2010)  127301}, \href{http://arxiv.org/abs/1005.3039}{{\tt arXiv:1005.3039 [astro-ph.CO]}}. [Erratum: Phys.Rev.D 82, 109902 (2010)].

\bibitem{Chen:2010va}
S.-H. Chen, J.~B. Dent, S.~Dutta, and E.~N. Saridakis, ``{Cosmological perturbations in f(T) gravity},'' \href{https://doi.org/10.1103/PhysRevD.83.023508}{{\em Phys. Rev. D} {\bf 83} (2011)  023508}, \href{http://arxiv.org/abs/1008.1250}{{\tt arXiv:1008.1250 [astro-ph.CO]}}.

\bibitem{Bahamonde:2019zea}
S.~Bahamonde, K.~Flathmann, and C.~Pfeifer, ``{Photon sphere and perihelion shift in weak $f(T)$ gravity},'' \href{https://doi.org/10.1103/PhysRevD.100.084064}{{\em Phys. Rev. D} {\bf 100} (2019) no.~8, 084064}, \href{http://arxiv.org/abs/1907.10858}{{\tt arXiv:1907.10858 [gr-qc]}}.

\bibitem{Paliathanasis:2017htk}
A.~Paliathanasis, J.~Levi~Said, and J.~D. Barrow, ``{Stability of the Kasner Universe in f(T) Gravity},'' \href{https://doi.org/10.1103/PhysRevD.97.044008}{{\em Phys. Rev. D} {\bf 97} (2018) no.~4, 044008}, \href{http://arxiv.org/abs/1709.03432}{{\tt arXiv:1709.03432 [gr-qc]}}.

\bibitem{Farrugia:2020fcu}
G.~Farrugia, J.~Levi~Said, and A.~Finch, ``{Gravitoelectromagnetism, Solar System Tests, and Weak-Field Solutions in $f (T,B)$ Gravity with Observational Constraints},'' \href{https://doi.org/10.3390/universe6020034}{{\em Universe} {\bf 6} (2020) no.~2, 34}, \href{http://arxiv.org/abs/2002.08183}{{\tt arXiv:2002.08183 [gr-qc]}}.

\bibitem{Bahamonde:2021srr}
S.~Bahamonde, A.~Golovnev, M.-J. Guzm\'an, J.~L. Said, and C.~Pfeifer, ``{Black holes in f(T,B) gravity: exact and perturbed solutions},'' \href{https://doi.org/10.1088/1475-7516/2022/01/037}{{\em JCAP} {\bf 01} (2022) no.~01, 037}, \href{http://arxiv.org/abs/2110.04087}{{\tt arXiv:2110.04087 [gr-qc]}}.

\bibitem{Bahamonde:2020bbc}
S.~Bahamonde, J.~Levi~Said, and M.~Zubair, ``{Solar system tests in modified teleparallel gravity},'' \href{https://doi.org/10.1088/1475-7516/2020/10/024}{{\em JCAP} {\bf 10} (2020)  024}, \href{http://arxiv.org/abs/2006.06750}{{\tt arXiv:2006.06750 [gr-qc]}}.

\bibitem{Bahamonde:2022ohm}
S.~Bahamonde, K.~F. Dialektopoulos, M.~Hohmann, J.~Levi~Said, C.~Pfeifer, and E.~N. Saridakis, ``{Perturbations in non-flat cosmology for f(T) gravity},'' \href{https://doi.org/10.1140/epjc/s10052-023-11322-3}{{\em Eur. Phys. J. C} {\bf 83} (2023) no.~3, 193}, \href{http://arxiv.org/abs/2203.00619}{{\tt arXiv:2203.00619 [gr-qc]}}.

\bibitem{Bahamonde:2020lsm}
S.~Bahamonde, V.~Gakis, S.~Kiorpelidi, T.~Koivisto, J.~Levi~Said, and E.~N. Saridakis, ``{Cosmological perturbations in modified teleparallel gravity models: Boundary term extension},'' \href{https://doi.org/10.1140/epjc/s10052-021-08833-2}{{\em Eur. Phys. J. C} {\bf 81} (2021) no.~1, 53}, \href{http://arxiv.org/abs/2009.02168}{{\tt arXiv:2009.02168 [gr-qc]}}.

\bibitem{Sotiriou:2008rp}
T.~P. Sotiriou and V.~Faraoni, ``{f(R) Theories Of Gravity},'' \href{https://doi.org/10.1103/RevModPhys.82.451}{{\em Rev. Mod. Phys.} {\bf 82} (2010)  451--497}, \href{http://arxiv.org/abs/0805.1726}{{\tt arXiv:0805.1726 [gr-qc]}}.

\bibitem{DeFelice:2010aj}
A.~De~Felice and S.~Tsujikawa, ``{f(R) theories},'' \href{https://doi.org/10.12942/lrr-2010-3}{{\em Living Rev. Rel.} {\bf 13} (2010)  3}, \href{http://arxiv.org/abs/1002.4928}{{\tt arXiv:1002.4928 [gr-qc]}}.

\bibitem{Bahamonde:2015zma}
S.~Bahamonde, C.~G. B\"ohmer, and M.~Wright, ``{Modified teleparallel theories of gravity},'' \href{https://doi.org/10.1103/PhysRevD.92.104042}{{\em Phys. Rev. D} {\bf 92} (2015) no.~10, 104042}, \href{http://arxiv.org/abs/1508.05120}{{\tt arXiv:1508.05120 [gr-qc]}}.

\bibitem{Capozziello:2018qcp}
S.~Capozziello, M.~Capriolo, and M.~Transirico, ``{The gravitational energy-momentum pseudotensor: the cases of $f(R)$ and $f(T)$ gravity},'' \href{https://doi.org/10.1142/S0219887818501645}{{\em Int. J. Geom. Meth. Mod. Phys.} {\bf 15} (2018)  1850164}, \href{http://arxiv.org/abs/1804.08530}{{\tt arXiv:1804.08530 [gr-qc]}}.

\bibitem{Bahamonde:2016grb}
S.~Bahamonde and S.~Capozziello, ``{Noether Symmetry Approach in $f(T,B)$ teleparallel cosmology},'' \href{https://doi.org/10.1140/epjc/s10052-017-4677-0}{{\em Eur. Phys. J. C} {\bf 77} (2017) no.~2, 107}, \href{http://arxiv.org/abs/1612.01299}{{\tt arXiv:1612.01299 [gr-qc]}}.

\bibitem{Farrugia:2018gyz}
G.~Farrugia, J.~Levi~Said, V.~Gakis, and E.~N. Saridakis, ``{Gravitational Waves in Modified Teleparallel Theories},'' \href{https://doi.org/10.1103/PhysRevD.97.124064}{{\em Phys. Rev. D} {\bf 97} (2018) no.~12, 124064}, \href{http://arxiv.org/abs/1804.07365}{{\tt arXiv:1804.07365 [gr-qc]}}.

\bibitem{Bahamonde:2016cul}
S.~Bahamonde, M.~Zubair, and G.~Abbas, ``{Thermodynamics and cosmological reconstruction in $f(T,B)$ gravity},'' \href{https://doi.org/10.1016/j.dark.2017.12.005}{{\em Phys. Dark Univ.} {\bf 19} (2018)  78--90}, \href{http://arxiv.org/abs/1609.08373}{{\tt arXiv:1609.08373 [gr-qc]}}.

\bibitem{Wright:2016ayu}
M.~Wright, ``{Conformal transformations in modified teleparallel theories of gravity revisited},'' \href{https://doi.org/10.1103/PhysRevD.93.103002}{{\em Phys. Rev. D} {\bf 93} (2016) no.~10, 103002}, \href{http://arxiv.org/abs/1602.05764}{{\tt arXiv:1602.05764 [gr-qc]}}.

\bibitem{Bahamonde:2019shr}
S.~Bahamonde, K.~F. Dialektopoulos, and J.~Levi~Said, ``{Can Horndeski Theory be recast using Teleparallel Gravity?},'' \href{https://doi.org/10.1103/PhysRevD.100.064018}{{\em Phys. Rev. D} {\bf 100} (2019) no.~6, 064018}, \href{http://arxiv.org/abs/1904.10791}{{\tt arXiv:1904.10791 [gr-qc]}}.

\bibitem{Bahamonde_2020}
S.~Bahamonde, K.~F. Dialektopoulos, V.~Gakis, and J.~L. Said, ``Reviving horndeski theory using teleparallel gravity after gw170817,'' \href{https://doi.org/10.1103/PhysRevD.101.084060}{{\em Physical Review D} {\bf 101} (2020) no.~8, 084060}. \url{http://dx.doi.org/10.1103/PhysRevD.101.084060}.

\bibitem{Bahamonde:2020cfv}
S.~Bahamonde, K.~F. Dialektopoulos, M.~Hohmann, and J.~Levi~Said, ``{Post-Newtonian limit of Teleparallel Horndeski gravity},'' \href{https://doi.org/10.1088/1361-6382/abc441}{{\em Class. Quant. Grav.} {\bf 38} (2020) no.~2, 025006}, \href{http://arxiv.org/abs/2003.11554}{{\tt arXiv:2003.11554 [gr-qc]}}.

\bibitem{Bernardo:2021qhu}
R.~C. Bernardo and J.~Levi~Said, ``{A data-driven reconstruction of Horndeski gravity via the Gaussian processes},'' \href{https://doi.org/10.1088/1475-7516/2021/09/014}{{\em JCAP} {\bf 09} (2021)  014}, \href{http://arxiv.org/abs/2105.12970}{{\tt arXiv:2105.12970 [astro-ph.CO]}}.

\bibitem{Bahamonde:2021dqn}
S.~Bahamonde, M.~Caruana, K.~F. Dialektopoulos, V.~Gakis, M.~Hohmann, J.~Levi~Said, E.~N. Saridakis, and J.~Sultana, ``{Gravitational-wave propagation and polarizations in the teleparallel analog of Horndeski gravity},'' \href{https://doi.org/10.1103/PhysRevD.104.084082}{{\em Phys. Rev. D} {\bf 104} (2021) no.~8, 084082}, \href{http://arxiv.org/abs/2105.13243}{{\tt arXiv:2105.13243 [gr-qc]}}.

\bibitem{Bernardo:2021bsg}
R.~C. Bernardo, J.~L. Said, M.~Caruana, and S.~Appleby, ``{Well-tempered Minkowski solutions in teleparallel Horndeski theory},'' \href{https://doi.org/10.1088/1361-6382/ac36e4}{{\em Class. Quant. Grav.} {\bf 39} (2022) no.~1, 015013}, \href{http://arxiv.org/abs/2108.02500}{{\tt arXiv:2108.02500 [gr-qc]}}.

\bibitem{Dialektopoulos:2021ryi}
K.~F. Dialektopoulos, J.~L. Said, and Z.~Oikonomopoulou, ``{Classification of teleparallel Horndeski cosmology via Noether symmetries},'' \href{https://doi.org/10.1140/epjc/s10052-022-10201-7}{{\em Eur. Phys. J. C} {\bf 82} (2022) no.~3, 259}, \href{http://arxiv.org/abs/2112.15045}{{\tt arXiv:2112.15045 [gr-qc]}}.

\bibitem{Capozziello:2023foy}
S.~Capozziello, M.~Caruana, J.~Levi~Said, and J.~Sultana, ``{Ghost and Laplacian instabilities in teleparallel Horndeski gravity},'' \href{https://doi.org/10.1088/1475-7516/2023/03/060}{{\em JCAP} {\bf 03} (2023)  060}, \href{http://arxiv.org/abs/2301.04457}{{\tt arXiv:2301.04457 [gr-qc]}}.

\bibitem{Ahmedov:2023num}
B.~Ahmedov, K.~F. Dialektopoulos, J.~Levi~Said, A.~Nosirov, Z.~Oikonomopoulou, and O.~Yunusov, ``{Cosmological perturbations in the teleparallel analog of Horndeski gravity},'' \href{https://doi.org/10.1088/1475-7516/2023/08/074}{{\em JCAP} {\bf 08} (2023)  074}, \href{http://arxiv.org/abs/2306.13473}{{\tt arXiv:2306.13473 [gr-qc]}}.

\bibitem{Ahmedov:2023lot}
B.~Ahmedov, K.~F. Dialektopoulos, J.~Levi~Said, A.~Nosirov, Z.~Oikonomopoulou, and O.~Yunusov, ``{Stable bouncing solutions in Teleparallel Horndeski gravity: violations of the no-go theorem},'' \href{http://arxiv.org/abs/2311.11977}{{\tt arXiv:2311.11977 [gr-qc]}}.

\bibitem{Ahmedov:2024aez}
B.~Ahmedov, M.~Caruana, K.~F. Dialektopoulos, J.~Levi~Said, A.~Nosirov, Z.~Oikonomopoulou, and O.~Yunusov, ``{Gauge invariant perturbations in teleparallel Horndeski gravity},'' \href{https://doi.org/10.1016/j.dark.2025.101846}{{\em Phys. Dark Univ.} {\bf 48} (2025)  101846}, \href{http://arxiv.org/abs/2412.01349}{{\tt arXiv:2412.01349 [gr-qc]}}.

\bibitem{Capozziello:2022zzh}
S.~Capozziello, V.~De~Falco, and C.~Ferrara, ``{Comparing equivalent gravities: common features and differences},'' \href{https://doi.org/10.1140/epjc/s10052-022-10823-x}{{\em Eur. Phys. J. C} {\bf 82} (2022) no.~10, 865}, \href{http://arxiv.org/abs/2208.03011}{{\tt arXiv:2208.03011 [gr-qc]}}.

\bibitem{Kofinas:2014daa}
G.~Kofinas and E.~N. Saridakis, ``{Cosmological applications of $F(T,T_G)$ gravity},'' \href{https://doi.org/10.1103/PhysRevD.90.084045}{{\em Phys. Rev. D} {\bf 90} (2014)  084045}, \href{http://arxiv.org/abs/1408.0107}{{\tt arXiv:1408.0107 [gr-qc]}}.

\bibitem{Kofinas:2014owa}
G.~Kofinas and E.~N. Saridakis, ``{Teleparallel equivalent of Gauss-Bonnet gravity and its modifications},'' \href{https://doi.org/10.1103/PhysRevD.90.084044}{{\em Phys. Rev. D} {\bf 90} (2014)  084044}, \href{http://arxiv.org/abs/1404.2249}{{\tt arXiv:1404.2249 [gr-qc]}}.

\bibitem{Kofinas:2014aka}
G.~Kofinas, G.~Leon, and E.~N. Saridakis, ``{Dynamical behavior in $f(T,T_G)$ cosmology},'' \href{https://doi.org/10.1088/0264-9381/31/17/175011}{{\em Class. Quant. Grav.} {\bf 31} (2014)  175011}, \href{http://arxiv.org/abs/1404.7100}{{\tt arXiv:1404.7100 [gr-qc]}}.

\bibitem{delaCruz-Dombriz:2017lvj}
A.~de~la Cruz-Dombriz, G.~Farrugia, J.~L. Said, and D.~Saez-Gomez, ``{Cosmological reconstructed solutions in extended teleparallel gravity theories with a teleparallel Gauss\textendash{}Bonnet term},'' \href{https://doi.org/10.1088/1361-6382/aa93c8}{{\em Class. Quant. Grav.} {\bf 34} (2017) no.~23, 235011}, \href{http://arxiv.org/abs/1705.03867}{{\tt arXiv:1705.03867 [gr-qc]}}.

\bibitem{delaCruz-Dombriz:2018nvt}
A.~de~la Cruz-Dombriz, G.~Farrugia, J.~L. Said, and D.~S\'aez-Chill\'on~G\'omez, ``{Cosmological bouncing solutions in extended teleparallel gravity theories},'' \href{https://doi.org/10.1103/PhysRevD.97.104040}{{\em Phys. Rev. D} {\bf 97} (2018) no.~10, 104040}, \href{http://arxiv.org/abs/1801.10085}{{\tt arXiv:1801.10085 [gr-qc]}}.

\bibitem{Kadam_2023}
S.~A. Kadam, B.~Mishra, and J.~Levi~Said, ``Noether symmetries in $f(t,t_g)$ cosmology,'' \href{https://doi.org/10.1088/1402-4896/acc0ac}{{\em Physica Scripta} {\bf 98} (2023)  }.

\bibitem{KADAM2024169563}
S.~Kadam, S.~V. Lohakare, and B.~Mishra, ``Dynamical complexity in teleparallel gauss bonnet gravity,'' \href{https://doi.org/https://doi.org/10.1016/j.aop.2023.169563}{{\em Annals of Physics} {\bf 460} (2024)  }.

\bibitem{LIGOScientific:2017vwq}
{\bf LIGO Scientific, Virgo} Collaboration, B.~P. Abbott {\em et al.}, ``{GW170817: Observation of Gravitational Waves from a Binary Neutron Star Inspiral},'' \href{https://doi.org/10.1103/PhysRevLett.119.161101}{{\em Phys. Rev. Lett.} {\bf 119} (2017) no.~16, 161101}, \href{http://arxiv.org/abs/1710.05832}{{\tt arXiv:1710.05832 [gr-qc]}}.

\bibitem{Goldstein:2017mmi}
A.~Goldstein {\em et al.}, ``{An Ordinary Short Gamma-Ray Burst with Extraordinary Implications: Fermi-GBM Detection of GRB 170817A},'' \href{https://doi.org/10.3847/2041-8213/aa8f41}{{\em Astrophys. J. Lett.} {\bf 848} (2017) no.~2, L14}, \href{http://arxiv.org/abs/1710.05446}{{\tt arXiv:1710.05446 [astro-ph.HE]}}.

\bibitem{Horndeski:1974wa}
G.~W. Horndeski, ``{Second-order scalar-tensor field equations in a four-dimensional space},'' \href{https://doi.org/10.1007/BF01807638}{{\em Int. J. Theor. Phys.} {\bf 10} (1974)  363--384}.

\bibitem{Kobayashi:2019hrl}
T.~Kobayashi, ``{Horndeski theory and beyond: a review},'' \href{https://doi.org/10.1088/1361-6633/ab2429}{{\em Rept. Prog. Phys.} {\bf 82} (2019) no.~8, 086901}, \href{http://arxiv.org/abs/1901.07183}{{\tt arXiv:1901.07183 [gr-qc]}}.

\bibitem{Ezquiaga_2017}
J.~M. Ezquiaga and M.~Zumalacárregui, ``Dark energy after gw170817: Dead ends and the road ahead,'' \href{https://doi.org/10.1103/PhysRevLett.119.251304}{{\em Physical Review Letters} {\bf 119} (2017) no.~25, 251304}. \url{http://dx.doi.org/10.1103/PhysRevLett.119.251304}.

\bibitem{Creminelli_2017}
P.~Creminelli and F.~Vernizzi, ``{Dark Energy after GW170817 and GRB170817A},'' \href{https://doi.org/10.1103/PhysRevLett.119.251302}{{\em Physical Review Letters} {\bf 119} (2017)  251302}.

\bibitem{Sakstein_2017}
J.~Sakstein and B.~Jain, ``Implications of the neutron star merger gw170817 for cosmological scalar-tensor theories,'' \href{https://doi.org/10.1103/PhysRevLett.119.251303}{{\em Physical Review Letters} {\bf 119} (2017) no.~25, }. \url{http://dx.doi.org/10.1103/PhysRevLett.119.251303}.

\bibitem{Baker_2017}
T.~Baker, E.~Bellini, P.~G. Ferreira, M.~Lagos, J.~Noller, and I.~Sawicki, ``Strong constraints on cosmological gravity from gw170817 and grb 170817a,'' \href{https://doi.org/10.1103/PhysRevLett.119.251301}{{\em Physical Review Letters} {\bf 119} (2017) no.~25, }. \url{http://dx.doi.org/10.1103/PhysRevLett.119.251301}.

\bibitem{Misner1973}
C.~W. Misner, K.~S. Thorne, and J.~A. Wheeler, {\em {Gravitation}}.
\newblock W. H. Freeman and Co., San Francisco, 1973.

\bibitem{aldrovandi2004spin}
R.~Aldrovandi, P.~B. Barros, and J.~G. Pereira, ``Spin and anholonomy in general relativity,'' \href{http://arxiv.org/abs/gr-qc/0402022}{{\tt arXiv:gr-qc/0402022 [gr-qc]}}. \url{https://arxiv.org/abs/gr-qc/0402022}.

\bibitem{Nojiri_2005}
S.~Nojiri and S.~D. Odintsov, ``Modified gauss-bonnet theory as gravitational alternative for dark energy,'' \href{https://doi.org/10.1016/j.physletb.2005.10.010}{{\em Physics Letters B} {\bf 631} (2005) no.~1-2, 1--6}. \url{http://dx.doi.org/10.1016/j.physletb.2005.10.010}.

\bibitem{DEFELICE20091}
A.~De~Felice and S.~Tsujikawa, ``Construction of cosmologically viable f(g) gravity models,'' \href{https://doi.org/10.1016/j.physletb.2009.03.060}{{\em Physics Letters B} {\bf 675} (2009) no.~1, 1--8}. \url{https://www.sciencedirect.com/science/article/pii/S0370269309003591}.

\bibitem{Bardeen1980}
J.~M. Bardeen, ``Gauge-invariant cosmological perturbations,'' \href{https://doi.org/10.1103/PhysRevD.22.1882}{{\em Phys. Rev. D} {\bf 22} (1980) no.~8, 1882--1905}.

\bibitem{heisenberg2023}
L.~Heisenberg and M.~Hohmann, ``{Gauge-invariant cosmological perturbations in general teleparallel gravity},'' \href{https://doi.org/10.1140/epjc/s10052-024-12810-w}{{\em Eur. Phys. J. C} {\bf 84} (2024)  }, \href{http://arxiv.org/abs/2311.05597}{{\tt arXiv:2311.05597 [gr-qc]}}.

\bibitem{Saltas_2014}
I.~D. Saltas, I.~Sawicki, L.~Amendola, and M.~Kunz, ``Anisotropic stress as a signature of nonstandard propagation of gravitational waves,'' \href{https://doi.org/10.1103/PhysRevLett.113.191101}{{\em Physical Review Letters} {\bf 113} (2014) no.~19, 191101}.

\bibitem{Belgacem_2019}
Belgacem {\em et al.}, ``Testing modified gravity at cosmological distances with lisa standard sirens,'' \href{https://doi.org/10.1088/1475-7516/2019/07/024}{{\em Journal of Cosmology and Astroparticle Physics} {\bf 2019} (2019)  }.

\bibitem{Cai_2018}
Y.-F. Cai, C.~Li, E.~N. Saridakis, and L.-Q. Xue, ``{$f(T)$} gravity after gw170817 and grb170817a,'' \href{https://doi.org/10.1103/PhysRevD.97.103513}{{\em Physical Review D} {\bf 97} (2018) no.~10, }. \url{http://dx.doi.org/10.1103/PhysRevD.97.103513}.

\bibitem{DeFelice:2010sh}
A.~De~Felice, J.-M. Gerard, and T.~Suyama, ``{Cosmological perturbation in f(R,G) theories with a perfect fluid},'' \href{https://doi.org/10.1103/PhysRevD.82.063526}{{\em Phys. Rev. D} {\bf 82} (2010)  063526}, \href{http://arxiv.org/abs/1005.1958}{{\tt arXiv:1005.1958 [astro-ph.CO]}}.

\bibitem{Nishizawa_2018}
A.~Nishizawa, ``Generalized framework for testing gravity with gravitational-wave propagation. i. formulation,'' \href{https://doi.org/10.1103/physrevd.97.104037}{{\em Physical Review D} {\bf 97} (2018)  }.

\end{thebibliography}\endgroup

\end{document}